\documentclass[11pt]{article}
\usepackage{axodraw,mathrsfs}

\makeatletter
\@addtoreset{equation}{section}
\makeatother

\renewcommand{\Re}{\mathop{\rm Re\,}\nolimits}
\renewcommand{\Im}{\mathop{\rm Im\,}\nolimits}
\newcommand{\Tr}{\mathop{\rm Tr}}
\renewcommand{\slash}[1]{\rlap/#1}
\newcommand{\scatt}[4]{\ell_#1 \ell_#2 \to \ell_#3 \ell_#4}
\newcommand{\bscatt}[4]{\bar\ell_#1 \bar\ell_#2 \to \bar\ell_#3
\bar\ell_#4} 
\def\momscatt#1(#2)#3(#4)#5(#6)#7(#8){\ell_#1(#2) \ell_#3(#4) \to
\ell_#5(#6) \ell_#7(#8)}
\newcommand{\mutoeee}{\mu\rightarrow e e \bar e}
\newcommand{\muetoee}{\mu e\rightarrow e e}
\newcommand{\bmutoeee}{\bar\mu\rightarrow \bar e \bar e e}

\newcommand{\ave}[1]{\left<#1\right>}

\title{
\centerline{\normalsize hep-ph/0210232 \hfill SINP/TNP/02-29}\bigskip
\bf CP-violating Majorana phases, lepton-conserving processes
and final state interactions}
 
\author{
\bf Jos\'e F. Nieves\\
Laboratory of Theoretical Physics, 
Department of Physics, P.O. Box 23343\\
University of Puerto Rico, R\'{\i}o Piedras,
Puerto Rico 00931-3343
\and
\bf Palash B. Pal\\
Saha Institute of Nuclear Physics, 1/AF Bidhan-Nagar, 
Calcutta 700064, India
}

\date{October 2002}

\begin{document}
\maketitle

\begin{abstract}
The $CP$-violating phases associated with Majorana neutrinos can give
rise to $CP$-violating effects even in processes that conserve total
Lepton number, such as $\mutoeee$, $\muetoee$ and others.  After
explaining the reasons that make this happen, we consider the
calculation of the rates for the process of the form $\scatt abac$ and its
conjugate $\bscatt abac$, where $\ell_a, \ell_b, \ell_c$ denote charged
leptons of different flavors.  In the context of the Standard Model
with Majorana neutrinos, we show that the difference in the rates
depends on such phases.  Our calculations illustrate in detail the
mechanics that operate behind the scene, and set the stage for
carrying out the analogous, more complicated (as we explain),
calculations for other processes such as $\mutoeee$ and its
conjugate.
\end{abstract}

\section{Introduction}
Sometime ago \cite{Nieves:1987pp} we introduced a prescription for
identifying a minimal set of $CP$-violating parameters in the lepton
sector that are invariant under a rephasing of the fermion fields.
The prescription holds for any number of fermion generations and, more
importantly, accommodate the case in which the neutrinos are Majorana
particles.  Recently \cite{Nieves:2001fc}, based on that work we
analyzed the dependence of the squared amplitudes on the
rephasing-invariants of the lepton sector with Majorana neutrinos, for
various lepton-violating or lepton-conserving processes, giving
special attention to the dependence on the extra $CP$-violating
parameters that are due to the Majorana nature of the
neutrinos~\cite{Bilenky:1980cx,Schechter:1980gr,Doi:1981yb}.

It was widely believed that these extra parameters appear only in
lepton number violating processes \cite{Schechter:1981gk}, and there
is a lot of discussion in the literature about their possible
observable effects%
\cite{O'Donnell:1995ak,Rangarajan:1998hj,%
Liu:2001xs,Bilenky:2001rz,Aguilar-Saavedra:2000vr,Pascoli:2001by}.
However, in our recent work \cite{Nieves:2001fc}, we showed that they
can appear in lepton-number conserving processes as well. The true
condition for the occurrence of these parameters in a given process
seems to be the violation of lepton number on any fermion line in the
corresponding diagrams, and not necessarily that total lepton number
be violated by the process as a whole.  In processes that conserve the
total lepton number, there are in general diagrams in which the
individual fermion lines change the lepton number, but do so in such a
way that the changes between different lines cancel in the overall
diagram.  The interference terms produced by such diagrams contain the
extra $CP$-violating parameters that exist due to the Majorana nature
of the neutrinos.

However, this analysis was carried out by considering various generic
physical processes, classified according to whether they conserve
total lepton number, or by how many units they violate it, and then
finding their generic dependence on the rephasing-invariant
$CP$-violating parameters. The question of what is the mechanics that
operates in a specific process to give rise to such effects was not
considered there. This question is important when we attempt to
consider the difference in the rates for, for example, $\mutoeee$ vs
$\bmutoeee$, due to the $CP$ violating phases.  The issue here is
that, unless the final-state interactions are taken into account, the
calculation of the two rates will be equal (by the $CPT$ theorem) in
spite of the fact that $CP$ may be violated. Thus, while the arguments
and analysis of Ref. \cite{Nieves:2001fc} are indicative, a specific
calculation of the effects depends in general on the kinematical and
dynamical aspects of the particular process considered.

In order to fill this gap, we consider in the present paper the
calculation of the difference of the rates for some leptonic processes
and their conjugate ones, in the context of the Standard Model with
Majorana neutrinos.  The processes that we consider have the virtue
that they have a two-body final state, which makes it simpler to take
into account the final-state interactions, yet they contain all the
ingredients to understand and illustrate the issues that we have
mentioned. In addition, some of the formulas that we will present on
the way, are also required ingredients in the corresponding
calculations for other processes such as $\mutoeee$.

In Section \ref{sec:general} we make some general remarks about the
type of process we consider, explain why the effect can be seen in
some of the processes and not in others, mention the need to consider
the effect of the final state interactions, and set the stage for the
calculations that make up the rest of the paper.  The effect of the
final state interactions, which show up as an absorptive term in the
total amplitude, is calculated in Section\ \ref{sec:finalstate}, for
both the direct process and its conjugate. Based on those results, in
Section\ \ref{sec:rates} we compute the difference of the differential
rates for the process and its conjugate, and thus we are able to show
explicitly that it is given in terms of the rephasing-invariant
$CP$-violating parameters for Majorana neutrinos. Our outlook and
conclusions are given in Section\ \ref{sec:conclusions}. Four
appendices contain some of the details of various stages of the
calculation, including a Fierz transformation formula used, the
Cutkosky rules employed to determine the absorptive term due to the
final state interactions, and the phase space integrals over the
intermediate states required to implement the Cutkosky formula.

\section{General Remarks}
\label{sec:general}
We want to consider processes of the generic form
\begin{eqnarray}
\label{generic}
\scatt abcd
\end{eqnarray}
where $a,b,c,d$ take values from anyone of the lepton flavors
$e,\mu,\tau$, with the condition that they are not all equal.  Such
processes conserve total lepton number but in general violate the
individual lepton flavors. By the usual substitution (crossing) rules,
our considerations also apply to those processes that are related to
these by crossing.

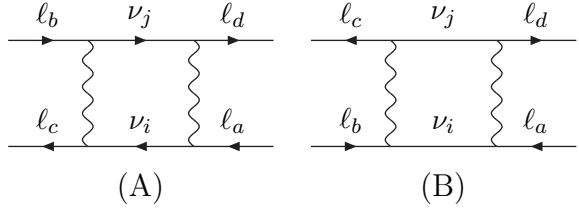
\begin{figure}
\begin{center}
\begin{picture}(100,70)(0,-15)
\ArrowLine(30,0)(0,0)
\Text(15,5)[b]{$\ell_c$}
\ArrowLine(70,0)(30,0)
\Text(50,5)[b]{$\nu_i$}
\ArrowLine(100,0)(70,0)
\Text(85,5)[b]{$\ell_a$}
\Photon(30,0)(30,40)24
\Photon(70,0)(70,40)24
\ArrowLine(0,40)(30,40)
\Text(15,45)[b]{$\ell_b$}
\ArrowLine(30,40)(70,40)
\Text(50,45)[b]{$\nu_j$}
\ArrowLine(70,40)(100,40)
\Text(85,45)[b]{$\ell_d$}
\Text(50,-10)[t]{\large (A)}
\end{picture}
\quad
\begin{picture}(100,70)(0,-15)
\ArrowLine(0,0)(30,0)
\Text(15,5)[b]{$\ell_b$}
\Line(30,0)(70,0)
\Text(50,5)[b]{$\nu_i$}
\ArrowLine(100,0)(70,0)
\Text(85,5)[b]{$\ell_a$}
\Photon(30,0)(30,40)24
\Photon(70,0)(70,40)24
\ArrowLine(30,40)(0,40)
\Text(15,45)[b]{$\ell_c$}
\Line(30,40)(70,40)
\Text(50,45)[b]{$\nu_j$}
\ArrowLine(70,40)(100,40)
\Text(85,45)[b]{$\ell_d$}
\Text(50,-10)[t]{\large (B)}
\end{picture}
\end{center}
\caption{The physical diagrams for the general process $\scatt abcd$. The
unlabeled vector boson lines represent the $W$ vector boson.  In
addition to these diagrams, there are other diagrams in which any one
of the $W$ bosons are replaced by their corresponding unphysical Higgs
particle, plus the exchange diagrams in which $\ell_c$ and $\ell_d$
interchanged.  Note that diagram (B) contributes only if the neutrinos
are Majorana particles.
\label{f:boxdiagrams}}
\end{figure}

We classify the processes into two groups, depending on whether or not
the diagrams that contain the one-loop photon or $Z$ vertex functions
also contribute. We denote by Group I the set of processes
for which the only diagrams that contribute are the box diagrams,
as shown in Fig.\ \ref{f:boxdiagrams}.
%
%
The requirement for the photon and $Z$ vertex diagrams to be absent
is that
\begin{eqnarray}
\label{cond1}
\ell_c \not = \ell_a \quad \mbox{and} \quad \ell_c \not = \ell_b 
\end{eqnarray}
and
\begin{eqnarray}
\label{cond2}
\ell_d \not = \ell_a \quad \mbox{and} \quad \ell_d \not = \ell_b 
\end{eqnarray}
Because each of the four indices $a,b,c,d$ can only take three
possible values ($e,\mu,\tau$), it is clear that the conditions in
(\ref{cond1}) and (\ref{cond2}) can be satisfied simultaneously only
if either
\begin{eqnarray}
\ell_a = \ell_b \,,
\end{eqnarray}
or if
\begin{eqnarray}
\ell_c = \ell_d \,.
\end{eqnarray}
Without any loss in generality, we can fix one or the other of the two
possibilities, and thus define the processes in Group I as those
of the general form
\begin{eqnarray}
\ell_a \ell_b\leftrightarrow \ell_c\ell_c 
\quad (\ell_a\not=\ell_b\not=\ell_c)\,.
\end{eqnarray}

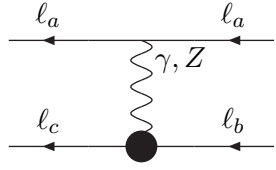
\begin{figure}
\begin{center}
\begin{picture}(100,70)(0,-15)
\ArrowLine(30,0)(0,0)
\Text(15,5)[b]{$\ell_c$}
\Line(30,0)(70,0)
\ArrowLine(100,0)(70,0)
\Text(85,5)[b]{$\ell_b$}
\Photon(50,0)(50,40){4}{4}
\Vertex(50,0){6}
\Text(65,28)[b]{$\gamma,Z$}
\ArrowLine(30,40)(0,40)
\Text(15,45)[b]{$\ell_a$}
\Line(30,40)(70,40)
\ArrowLine(100,40)(70,40)
\Text(85,45)[b]{$\ell_a$}
\end{picture}
\end{center}
\caption{Representation of the collection of diagrams
for the processes in Group II, that involve the one-loop
$\gamma\ell_b\ell_c$ or $Z\ell_b\ell_c$ vertex functions.
\label{f:gammaZdiagrams}}
\end{figure}

On the other hand, the processes in Group II are those for which
one (or more) of the conditions in (\ref{cond1}) and (\ref{cond2})
is not satisfied. Again, without loss of generality, they can be
represented by the general form of Eq.\ (\ref{generic}) with
$\ell_d=\ell_a$, i.e., these processes have the generic form
\begin{eqnarray}
\label{genericII}
\scatt abac \,,
\end{eqnarray}
with $c \not= b$ being the only restriction.  In this case, in
addition to the diagrams shown in Fig.\ \ref{f:boxdiagrams}, the
diagrams shown schematically in Fig.\ \ref{f:gammaZdiagrams} must be
included.
%
%

\subsection{No-go result for Group I}
As already stated, the diagrams that contribute to the amplitude for
these processes are the box diagrams shown in
Fig.~\ref{f:boxdiagrams}. To the leading order ($1/M_W^4$), only the
diagrams involving the $W$ exchange are important. Let us consider
$\scatt abcc$.  The physical amplitude is of the form
\begin{eqnarray}
\label{AgenericI}
A\Big(\momscatt a(k)b(p)c(p_1)c(p_2)\Big) & = & M_A \Big(\momscatt a(k)b(p)c(p_1)c(p_2)\Big) \nonumber\\* 
&&\mbox{} + M_B \Big(\momscatt a(k)b(p)c(p_1)c(p_2)\Big)
\end{eqnarray}
where $M_X(\momscatt a(k)b(p)c(p_1)c(p_2))$ is the contribution from each diagram. 
Their calculation is straightforward, and some of the details
are provided in Appendix\ \ref{sec:calcboxdiagrams}.
The results for the present case are summarized by the formulas
\begin{eqnarray}
\label{MAB}
M_A \Big(\momscatt a(k)b(p)c(p_1)c(p_2)\Big) & = & \lambda^{(abcc)}_A 
[\bar u_c(p_2)\gamma^\mu L u_b(p)][\bar u_c(p_1)\gamma_\mu L u_a(k)] \,,
\nonumber\\
M_B \Big(\momscatt a(k)b(p)c(p_1)c(p_2) \Big) & = & \lambda^{(abcc)}_B 
[\bar u_c(p_2)\gamma^\mu L u_b(p)][\bar u_c(p_1)\gamma_\mu L u_a(k)] \,,
\end{eqnarray}
where
\begin{eqnarray}
\label{lambdaAB}
\lambda^{(abcd)}_A & \equiv & 
\frac{g^4}{64\pi^2 M_W^2} 
\sum_{i,j} (V^*_{a i}V^*_{b j}V_{c i}V_{d j})(4f^{(ij)}_A) +
(c\leftrightarrow d)\,, \nonumber\\
\lambda^{(abcd)}_B & \equiv & 
\frac{g^4}{64\pi^2 M_W^2} 
\sum_{i,j}\left(K_i^2 K_j^{*\,2} 
V_{ai}^* V_{bi}^* V_{cj} V_{dj}\right)(2f^{(ij)}_B) +
(c\leftrightarrow d) \,.
\end{eqnarray}
The functions $f^{(ij)}_A$ and $f^{(ij)}_B$ appearing in these
equations are given by
\begin{eqnarray}
\label{fAB}
f^{(ij)}_A & = & 
\frac{r^2_i \log r_i}{r_j - r_i} - r_i + (i\leftrightarrow j)\,,
\nonumber\\ 
f^{(ij)}_B & = &
\frac{m_{\nu_i}m_{\nu_j}}{M_W^2} \left\{
\frac{r_i \log r_i}{r_j - r_i} - r_i + (i\leftrightarrow j)\right\} \,,
\end{eqnarray}
where
\begin{eqnarray}
r_i = {m_{\nu_i}^2 \over M_W^2} \,.
\end{eqnarray}
Further, the $V_{a i}$ are the elements of the lepton mixing matrix,
and the $K_i$ are the phases defined by the Majorana condition
\begin{eqnarray}
\nu^c_i = K_i^2\nu_i \,.
\label{Majocond}
\end{eqnarray}
Although there are several equivalent ways to write the results for
the amplitudes $M_{A,B}$, they can be brought to this form by suitable
Fierz transformations.

The upshot of this is that when Eq.\ (\ref{MAB}) is substituted in
Eq.\ (\ref{AgenericI}), the effective couplings $\lambda_{A,B}$ appear
in combination as a common overall factor of the total amplitude.
Since there is no interference term, the rate for the process and its
conjugate is the same, and there is no observable $CP$ violating
effect in this type of process. We have considered explicitly the
amplitude for $\scatt abcc$, but similar arguments hold for the
inverse $\scatt ccab$, whose amplitude is simply the
complex conjugate, and other related processes such as
$\ell_a\to\bar\ell_b\ell_c\ell_c$.

While we have summarized the result of the actual calculation, a
little thought reveals what is going on. Because we are calculating to
leading order in $1/M_W^2$, the dominant terms come from the
$W$-exchange diagrams, as we have already mentioned. The chiral nature
of the $W$ interactions dictate that, to leading order, only the
left-handed components of the external fermion fields enter in the
amplitude.  The most economical way to express this fact is by writing
down the effective Lagrangian for this process which, by the above
argument, can only be of the form
\begin{eqnarray}
\label{LW}
\mathscr L^{(W)} = \frac{\lambda}{2}[\bar \ell_c\gamma^\mu L \ell_b]
[\bar \ell_c\gamma_\mu L \ell_a] + \mbox{h.c.}
\end{eqnarray}
In fact, the results given in Eqs.\ (\ref{AgenericI}) and (\ref{MAB})
can be represented by this Lagrangian, with the identification
$\lambda = \lambda^{(abcc)}_A + \lambda^{(abcc)}_B$. 
Thus, to this order the effective
Lagrangian actually consists of only one term, and therefore the rates for
the process and its conjugate are equal.

\subsection{Evasion for Group II}
By the same argument, it is now easy to see how the processes in Group
II differ.  We consider specifically those with $\ell_a \not=
\ell_b$.  The diagrams for a process of the generic form given in
Eq.\ (\ref{genericII}), include the diagrams that involve the
$\gamma\ell_b\ell_c$ and $Z\ell_b\ell_c$ one-loop vertex
functions \cite{Petcov:1976ff}.  
Instead of Eq.\ (\ref{AgenericI}), the physical amplitude
in this case is of the form
\begin{eqnarray}
\label{AgenericII}
A\Big(\momscatt a(k)b(p)a(k')c(p')\Big) 
& = & (\lambda^{(abac)}_A + \lambda^{(abac)}_B)
[\bar u_a(k')\gamma^\mu L u_a(k)]
[\bar u_c(p')\gamma_\mu L u_b(p)]\nonumber\\*[8pt]
&&\mbox{} + \lambda^{(bc)}_Z
[\bar u_a(k')\gamma^\mu(X + Y\gamma_5)u_a(k)]
[\bar u_c(p') \gamma_\mu L u_b(p)]\,,\nonumber\\*
\end{eqnarray}
where
\begin{eqnarray}
\label{lambdaZ}
\lambda^{(bc)}_Z = - \frac{g^4}{64\pi^2 M_W^2}
\sum_k V^*_{b k} V_{c k} f^{(k)}_Z \,,
\end{eqnarray}
with
\begin{eqnarray}
\label{fZ}
f^{(k)}_Z = r_k\log r_k \,,
\end{eqnarray}
while $X$ and $Y$ are the neutral-current couplings
of the lepton $\ell_a$, 
\begin{eqnarray}
\label{XaYa}
X & = & -\frac{1}{2} + \sin^2\theta_W \nonumber\\* 
Y & = & \frac{1}{2} \,.
\end{eqnarray}
We mention the following. In the formula quoted in Eq.\ (\ref{lambdaZ})
we have neglected the other terms of order $1/M_W^4$ that do not
contain the logarithmic factor $\log r_i$. In addition, none of the terms
that arise from the diagram that involve the photon vertex function
contain that logarithmic factor, and therefore we have omitted altogether
that contribution in Eq.\ (\ref{AgenericII}).
Thus, by the same argument that led us to write Eq.\ (\ref{LW}),
this amplitude corresponds to an effective Lagrangian
\begin{eqnarray}
\label{LWZ}
\mathscr L^{(W+Z)} &=& 
(\lambda^{(abac)}_A + \lambda^{(abac)}_B) 
[\bar \ell_a\gamma^\mu L \ell_a]
[\bar \ell_c\gamma_\mu L \ell_b] \nonumber\\*
&& +
\lambda^{(bc)}_Z
[\bar \ell_a\gamma^\mu(X + Y\gamma_5)\ell_a]
[\bar \ell_c \gamma_\mu L \ell_b] + \mbox{h.c.} \,.
\end{eqnarray}
This in turn can be written in the more compact form
\begin{eqnarray}
\label{LWZcompact}
\mathscr L^{(W+Z)}  = 
[\bar \ell_a\gamma^\mu(X' + Y'\gamma_5)\ell_a]
[\bar \ell_c \gamma_\mu L \ell_b] + h.c. \,,
\end{eqnarray}
with
\begin{eqnarray}
\label{XYprime}
X' & = & \lambda^{(abac)}_A + \lambda^{(abac)}_B + \lambda^{(bc)}_Z X
\nonumber\\*
Y' & = & -(\lambda^{(abac)}_A + \lambda^{(abac)}_B) + \lambda^{(bc)}_Z
Y \,. 
\end{eqnarray}
In particular, $X'$ and $Y'$ are complex quantities, and depend on the
Majorana phases through the $\lambda_B$ term.

A kinematic observable that depends on the interference term 
$\Im(X'^* Y')$ will be sensitive to the $CP$-violating
Majorana phases. However, the total rate, determined from this
effective Lagrangian, or equivalently calculated with the amplitude
given in Eq.\ (\ref{AgenericII}), will not depend on that interference
term. The reason is that, as seen from Eq.\ (\ref{LWZcompact}), such
term arises from the interference between the vector and axial vector
parts of the current, and that vanishes after summing and averaging
over the polarizations and integrating over the phase space.
Moreover, the rates for any process and its conjugate, determined from
Eq.\ (\ref{LWZcompact}), are equal. This is ultimately due to the fact
that, while $CP$ does not hold in Eq.\ (\ref{LWZcompact}), $CPT$ does
hold and that is sufficient to guarantee the equality of the rates.

When we consider the final state interactions between the outgoing
leptons, this is no longer true as is well known. While the amplitude
for the direct process depends on $X'$ and $Y'$, the
amplitude for the conjugate process depends on the complex conjugates
of these two quantities. However, the final state interactions induce
an extra phase that is the same for both the direct process and the
conjugate. This mismatch between the two sets of phases in both cases
leads to the inequality of the total rates. In the language of the
effective Lagrangian, the effect of the final state interactions is to
augment Eq.\ (\ref{LWZcompact}) in a way that renders it non-hermitian.
Our task in the next section is to calculate the effect that we have 
just outlined. 

We would like to remark that the above statements about the equality
or inequality of the total rates, as the case may be, apply to the
total differential rates as well. The latter quantities are defined
from the squared amplitude as usual, by summing over the final spins
and averaging over the initial ones, with the total integrated rates
being obtained from them by carrying out the only non-trivial
integration, over the azimuthal angle.  The reason why we can also
consider the total differential rates is the following. Under a $CP$
or $CPT$ transformation, the amplitude is related to the amplitude for
the conjugate process, with perhaps the momentum and/or spin variables
reversed. For the total integrated rates, the reversal of the momentum
and spin variables is of no consequence since they are being summed
over. In the case of the total differential rates, the reversal of the
spin variables is not relevant either for the same reason.  On the
other hand, for the two body processes that we are considering, the
momentum vectors can appear only through the scalar variables formed
out of the scalar products among them, and those variables are
unchanged by the simultaneous transformations of the momentum
vectors\footnote{The situation is different if we consider for
example, processes with three particles in the final state. In that
case, the spin-averaged squared amplitude can contain momentum
contractions involving the four-dimensional antisymmetric tensor, and
those change when the momentum vectors are reversed.  Our
considerations could be applied to such cases also, but only if some
(non-trivial) angular integrations are made so that those terms do not
appear in some specially designed differential rates.}.

\section{Final state interactions}
\label{sec:finalstate}
\subsection{General considerations}
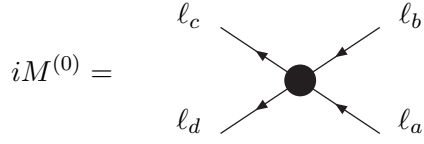
\begin{figure}
\begin{center}
\begin{picture}(100,70)(-20,-15)
\Text(-60,20)[b]{$iM^{(0)}  =$}
\ArrowLine(30,20)(0,0)
\ArrowLine(30,20)(0,40)
\ArrowLine(60,40)(30,20)
\ArrowLine(60,0)(30,20)
\Vertex(30,20){6}
\Text(-12,0)[b]{$\ell_d$}
\Text(72,0)[b]{$\ell_a$}
\Text(-12,40)[b]{$\ell_c$}
\Text(72,40)[b]{$\ell_b$}
\end{picture}
\end{center}
\caption{Schematic representation of the one-loop amplitude for
$\scatt abcd$. The diagram stands for the collection of diagrams referred
to in Figs.\ \ref{f:boxdiagrams} and \ref{f:gammaZdiagrams}.
\label{f:1-loop}}
\end{figure}

\begin{figure}
\begin{center}
\begin{picture}(100,70)(-20,-15)
\Text(-60,20)[b]{$iM^{(1)}=\null$}
\ArrowLine(30,20)(0,0)
\ArrowLine(30,20)(0,40)
\ArrowLine(60,40)(30,20)
\ArrowLine(60,0)(30,20)
\Vertex(30,20){6}
\Text(-12,0)[b]{$\ell_d$}
\Text(72,0)[b]{$\ell_a$}
\Text(-12,40)[b]{$\ell_c$}
\Text(72,40)[b]{$\ell_b$}
\Photon(9,6)(9,34)24
\Text(0,21)[c]{$\gamma$}
\end{picture}
\end{center}
\caption{Schematic representation of the correction
to include the electromagnetic final state interactions in $\scatt
abcd$.\label{f:2-loop}}
\end{figure}
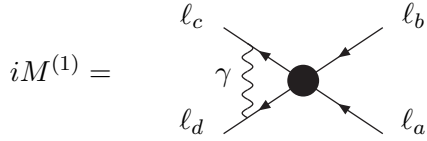

We denote by $M^{(0)}(\scatt abac)$ the amplitude for $\scatt abac$
determined from Eq.\ (\ref{LWZcompact}), 
which we represent schematically in Fig.\ \ref{f:1-loop}.
Analogously, we denote by $M^{(0)}(\bscatt abac)$ the amplitude for the 
conjugate process.
%
%
%
As we have already mentioned, since the individual terms that
contribute to $M^{(0)}$ do not contain an absorptive part, then the
rates $\Gamma(\scatt abac)$ and $\Gamma(\bscatt abac)$ calculated with the
above amplitudes are equal, as a consequence of $CPT$.  This result
does not hold if we include the final state interactions between the
two leptons, of which the dominant one is the electromagnetic
interaction. Diagrammatically, the additional terms are represented in
Fig.\ \ref{f:2-loop}.  The total amplitude is
\begin{equation}
M = M^{(0)} + M^{(1)} \,.
\end{equation}
Using the Cutkosky rules, the absorptive part of $M^{(1)}$ can be
expressed in terms of essentially $M^{(0)}$ itself, times some
factors.
%
%

In order to write the following formulas in a compact form,
let us introduce the following notation. We denote the initial
and the final states in the process $\scatt abac$ by
\begin{eqnarray}
\left| i \right> & \equiv & \left|\ell_a(k,s_a)
\ell_b(p,s_b)\right> \nonumber\\*
\left|f\right> & \equiv & \left|\ell_a(k',s_a') \ell_c(p',s_c')
\right> \,.
\end{eqnarray}
The $S$-matrix element for this process is then written as
\begin{equation}
\label{Smatrix}
\left< f\left|S\right|i\right> = (2\pi)^4
\delta^{(4)}(k + p - k' - p')
\Big[ iM^{(0)}(\scatt abac) + iM^{(1)}(\scatt abac) \Big] \,,
\end{equation}
where $M^{(0)}(\scatt abac)$ is determined from Eq.\ (\ref{LWZcompact}).
Our next task is to find the appropriate expression for the two-loop
amplitude $M^{(1)}(\scatt abac)$.

In general, it can be decomposed as
\begin{equation}
M^{(1)}(\scatt abac) = M^{(1)}_{\rm disp}(\scatt abac) + 
M^{(1)}_{\rm abs}(\scatt abac) \,,
\end{equation}
where $M^{(1)}_{\rm disp}$ and $M^{(1)}_{\rm abs}$ stand for its
dispersive and absorptive parts, respectively.  Since, as already
mentioned, $M^{(0)}$ does not contain an absorptive part, the
contribution from $M^{(1)}_{\rm disp}$ only produces an order $\alpha$
correction to $M^{(0)}$ which we neglect.  On the other hand, although
$M^{(1)}_{\rm abs}$ is also a factor of order $\alpha$ smaller than
$M^{(0)}$, it is the piece we are after since it contributes an
absorptive part to the full amplitude.  To calculate it, we employ the
Cutkosky rules \cite{iz}. As shown in Appendix \ref{sec:cutkosky}, in
our notation for the present case they yield
\begin{equation}
\label{M1a}
M^{(1)}_{\rm abs}(\scatt abac) =
\frac{i}{2}\sum_{n} (2\pi)^4\delta^{(4)}(q_1 + q_2 - k' - p') 
M^{(\gamma)} \Big(\scatt {a^{(n)}} {c^{(n)}} ac \Big) 
M^{(0)} \Big(\scatt ab {a^{(n)}} {c^{(n)}} \Big) \,.
\end{equation}
Here the symbols $\ell^{(n)}_a \ell^{(n)}_c$ 
stand for an intermediate $\ell_a\ell_c$ state
\begin{equation}
\left|n\right> \equiv 
\left|\ell_a(q_1,\sigma_1) \ell_c(q_2,\sigma_2)\right> \,,
\end{equation}
and the sum over the intermediate states stands for
\begin{equation}
\sum_{n} \rightarrow \int\frac{d^3q_1}{(2\pi)^3 2 E^{(a)}_1}
\frac{d^3q_2}{(2\pi)^3 2E^{(c)}_2}\sum_{\sigma_1\,,\sigma_2} \,.
\end{equation}
The quantity $M^{(\gamma)}$ in Eq.\ (\ref{M1a}) is the electromagnetic
scattering amplitude for $\ell_a(q_1,\sigma_1) + \ell_c(q_2,\sigma_2) 
\rightarrow \ell_a(k',s_a') + \ell_c(p',s_c')$, i.e.,
\begin{eqnarray}
\label{Mgamma}
iM^{(\gamma)}\Big(\scatt {a^{(n)}} {c^{(n)}} ac \Big) =
\frac{ie^2}{(q_1 - k')^2}\left[\bar u_a(k',s_a')\gamma^\mu
u_a(q_1,\sigma_1)\right] 
\left[\bar u_c(p',s_c')\gamma_\mu u_c(q_2,\sigma_2)\right] \,.
\end{eqnarray}
Thus finally, the full amplitude is given by
\begin{equation}
\label{fullM}
M(\scatt abac) = M^{(0)}(\scatt abac) + M^{(1)}_{\rm abs}(\scatt abac)\,,
\end{equation}
where $M^{(1)}_{\rm abs}(\scatt abac)$ is computed
from Eq.\ (\ref{M1a}), with $M^{(0)}(\scatt abac)$ determined
from (\ref{LWZcompact}).

In a similar fashion, for the conjugate process $\bscatt abac$, we
define 
\begin{eqnarray}
\left|\bar i\right> & \equiv & \left|\bar\ell_a(k,s_a) 
\bar\ell_b(p,s_b)\right>\nonumber\\
\left|\bar f\right> & \equiv & 
\left|\bar\ell_a(k',s_a') \bar\ell_c(p',s_c')\right> \nonumber\\
\left|\bar n\right> & \equiv  &
\left|\bar\ell_a(q_1,\sigma_1) \bar\ell_c(q_2,\sigma_2)\right> \,.
\end{eqnarray}
Neglecting again the dispersive part of the amplitude,
the corresponding $S$-matrix element is given by
\begin{equation}
\label{Smatrixbar}
\left< \bar f\left|S\right|\bar i\right> = 
(2\pi)^4\delta^{(4)}(k + p - k' - p') \Big[
iM^{(0)}(\bscatt abac) + iM^{(1)}_{\rm abs}(\bscatt abac) \Big] \,.
\end{equation}
where
\begin{equation}
\label{M1bara}
M^{(1)}_{\rm abs}(\bscatt abac) = 
\frac{i}{2}\sum_{\bar n} (2\pi)^4\delta^{(4)}(q_1 + q_2 - k' - p') 
M^{(\gamma)} \Big(\bscatt {a^{(n)}} {c^{(n)}} ac \Big) 
M^{(0)}\Big(\bscatt ab {a^{(n)}} {c^{(n)}} \Big) \,,
\end{equation}
with
\begin{eqnarray}
iM^{(\gamma)} \Big(\bscatt {a^{(n)}} {c^{(n)}} ac \Big) 
= \frac{ie^2}{(q_1 - k')^2}
\left[\bar v_a(q_1,\sigma_1)\gamma^\mu
v_a(k',s_a')\right]
\left[\bar v_c(q_2,\sigma_2)\gamma_\mu
v_c(p',s_c')\right]\,.
\end{eqnarray}

It is useful to note the following.
Using the relation between the spinors $u(p,s)$ and $v(p,s)$,
e.g., 
\begin{equation}
\label{ccspinor}
v(p,s) = i\gamma_2 u^*(p,s) \,,
\end{equation}
in a specific convention, together with relations such as
\begin{equation}
\label{ccvector}
\bar v'\gamma_\mu v = \bar u\gamma_\mu u' \,,
\end{equation}
it follows that
\begin{equation}
\label{Mgammabar}
M^{(\gamma)} \Big(\bscatt {a^{(n)}} {c^{(n)}} ac \Big) = 
M^{(\gamma)} \Big(\scatt {a^{(n)}} {c^{(n)}} ac \Big) \,,
\end{equation}
that is, the electromagnetic amplitude for the two processes
is the same. 

Our task at hand is to apply these formulas to compute the absorptive
part of the amplitudes for the direct process and its conjugate, from
Eqs.\ (\ref{M1a}) and (\ref{M1bara}), respectively.

\subsection{The absorptive part $M^{(1)}_{\rm abs}(\scatt abac)$} 
First of all, from Eq.\ (\ref{LWZcompact}) we write
\begin{eqnarray}
M^{(0)} \Big(\scatt ab {a^{(n)}} {c^{(n)}} \Big) =
[\bar u_a(q_1)\gamma^\mu(X' + Y'\gamma_5) u_a(k)]
[\bar u_c(q_2) \gamma_\mu L u_b(p)] \,.
\end{eqnarray}
Using this and Eq.\ (\ref{Mgamma}), we then obtain
\begin{eqnarray}
M^{(1)}_{\rm abs}(\scatt abac) & = & 
\frac{ie^2}{2} \int dL
[\bar u_a(k')\gamma^\lambda(\slash q_1 + m_a)
\gamma^\mu(X' + Y'\gamma_5)u_a(k)]
\nonumber\\*
&&\times
[\bar u_c(p')\gamma_\lambda(\slash q_2 + m_c)
\gamma_\mu L u_b(p)] \,,
\end{eqnarray}
where the symbol $\displaystyle\int dL$ stands for
\begin{eqnarray}
\int dL \rightarrow
\int\frac{d^3q_1}{(2\pi)^3 2E^{(a)}_{q_1}}
\frac{d^3q_2}{(2\pi)^3 2E^{(c)}_{q_2}}
(2\pi)^4\delta^{(4)}(k' + p' - q_1 - q_2)
\frac{1}{(k' - q_1)^2} \,.
\label{dX}
\end{eqnarray}
In order to proceed, we introduce the following definitions for the
integrals over the intermediate momenta
\begin{eqnarray}
I^{(0)} & = & \int dL \nonumber\\
I^{(1)}_\mu & = & \int dL\; q_{1\mu} \nonumber\\
I^{(2)}_\mu & = & \int dL\; q_{2\mu} \nonumber\\
I^{(12)}_{\mu\nu} & = & \int dL\; q_{1\mu} q_{2\nu} \,,
\label{I...}
\end{eqnarray}
in terms of which
\begin{eqnarray}
M^{(1)}_{\rm abs}(\scatt abac) = \frac{ie^2}{2} \bigg( M^{(1)}_1 +
M^{(1)}_2 + M^{(1)}_3 + M^{(1)}_4 \bigg) \,,
\label{M1M2M3M4}
\end{eqnarray}
where
\begin{eqnarray}
\label{M1parts}
M^{(1)}_1 & = & 
m_a m_c
I^{(0)}\Big[ \bar u_a(k')\gamma^\lambda
\gamma^\mu(X' + Y'\gamma_5)u_a(k) \Big]
\Big[ \bar u_c(p')\gamma_\lambda \gamma_\mu L u_b(p) \Big]
\nonumber\\
M^{(1)}_2 & = & 
m_a I^{(2)}_\rho \Big[ \bar u_a(k')\gamma^\lambda
\gamma^\mu(X' + Y'\gamma_5)u_a(k) \Big]
\Big[ \bar u_c(p')\gamma_\lambda \gamma^\rho
\gamma_\mu L u_b(p) \Big]\nonumber\\
M^{(1)}_3 & = & 
m_c I^{(1)}_\rho \Big[ \bar u_a(k')\gamma^\lambda \gamma^\rho
\gamma^\mu(X' + Y'\gamma_5)u_a(k) \Big]
\Big[ \bar u_c(p')\gamma_\lambda\gamma_\mu L u_b(p) \Big]
\nonumber\\
M^{(1)}_4 & = & 
I^{(12)}_{\rho\nu}
\Big[ \bar u_a(k')\gamma^\lambda\gamma^\rho
\gamma^\mu(X' + Y'\gamma_5)u_a(k) \Big]
\Big[ \bar u_c(p')\gamma_\lambda \gamma^\nu \gamma_\mu L u_b(p) \Big] \,.
\end{eqnarray}
While the integrals are doable in the general case, the procedure is
tedious and the final formulas are cumbersome.  Therefore, for
the moment we proceed as far as possible without using the explicit
results of their evaluation.

\subsection{The absorptive part $M^{(1)}_{\rm abs}(\bscatt abac)$} 
We carry out the same procedure with the amplitude for the conjugate
process. From Eq.\ (\ref{LWZcompact}),
\begin{eqnarray}
M^{(0)} \Big(\bscatt ab {a^{(n)}} {c^{(n)}} \Big) =
\Big[ \bar v_a(k)\gamma^\mu(X'^* + 
Y'^*\gamma_5) v_a(q_1) \Big]
\Big[ \bar v_b(p) \gamma_\mu L v_\gamma(q_2) \Big] 
\end{eqnarray}
which, using relations such as those give in Eqs.\ (\ref{ccvector})
and (\ref{ccspinor}), can be written in the form
\begin{eqnarray}
M^{(0)} \Big(\bscatt ab {a^{(n)}} {c^{(n)}} \Big) =
\Big[ \bar u_a(q_1)\gamma^\mu(X'^* -
Y'^*\gamma_5) u_a(k) \Big]
\Big[ \bar u_c(q_2) \gamma_\mu R u_b(p) \Big] \,.
\end{eqnarray}
{From} Eq.\ (\ref{M1bara}), and using Eq.\ (\ref{Mgammabar}), 
we then obtain
\begin{eqnarray}
M^{(1)}_{\rm abs}(\bscatt abac) & = & \frac{ie^2}{2} \int dL
\Big[ \bar u_a(k')\gamma^\lambda(\slash q_1 + m_a)
\gamma^\mu(X'^* - Y'^*\gamma_5)u_a(k) \Big]
\nonumber\\*
&&\times
\Big[ \bar u_c(p')\gamma_\lambda(\slash q_2 + m_c)
\gamma_\mu R u_b(p) \Big] \,.
\end{eqnarray}
By comparison, it is immediately seen that the amplitude for this
process is obtained from the formulas for the direct process by making
the substitutions
\begin{eqnarray}
\label{M->Mbar}
X' & \rightarrow & X'^* \nonumber\\*
Y' & \rightarrow & Y'^* \nonumber\\*
\gamma_5 & \rightarrow & -\gamma_5 \,.
\end{eqnarray}
%

%
%
\section{The difference in the rates}
\label{sec:rates}
If we write the total amplitude in the form
\begin{eqnarray}
M & = & M^{(0)}(\scatt abac) + M^{(1)}_{\rm abs}(\scatt abac)
\nonumber\\*
\overline M & = & M^{(0)}(\bscatt abac) + M^{(1)}_{\rm abs}(\bscatt abac)\,,
\end{eqnarray}
the quantity in which we are interested is the difference
\begin{eqnarray}
\label{ratediffdefn}
\ave{|M|^2} - \ave{|\overline M|^2} & = &  
2\Re\ave{M^{(0)*}(\scatt abac)M^{(1)}_{\rm abs}(\scatt abac)} \nonumber\\*
& - &
2\Re\ave{M^{(0)*}(\bscatt abac)M^{(1)}_{\rm abs}(\bscatt abac)}\,,
\end{eqnarray}
where the angle bracket notation indicate the operation to sum and
average over the final and initial spins, respectively.  In this
expression we have made use of the fact that, after that operation is
made, the terms without the absorptive part cancel out.  Using Eq.\
(\ref{M1M2M3M4}), and denoting $M^{(0)}(\scatt abac)$ by simply
$M^{(0)}$, we can write
\begin{eqnarray}
\ave{|M|^2} - \ave{|\overline M|^2} & = &  R - \bar R \,,
\end{eqnarray}
where
\begin{eqnarray}
\label{Rdefn}
R \equiv -e^2 \sum_{i = 1}^4 \Im\ave{M^{(0)*}M^{(1)}_i}
\end{eqnarray}
and $\bar R$ is obtained from $R$ by making the substitutions
indicated in Eq.\ (\ref{M->Mbar}). We now compute the various terms
in $R$.

\subsection{1st term}
Averaging over initial spins and summing over final spins, we obtain
\begin{eqnarray}
\left< M^{(0)*} M^{(1)}_1 \right> &=& 
\frac14 m_a m_c I^{(0)}
\Tr\Big[(\slash p + m_b) \gamma_\alpha L (\slash p' + m_c)
\gamma_\lambda\gamma_\mu L \Big] \nonumber\\*
&\times& \Tr\Big[ (\slash k + m_a) \gamma^\alpha 
(X'^* + Y'^*\gamma_5)
(\slash k' + m_a) \gamma^\lambda \gamma^\mu (X' + Y'\gamma_5) \Big]\\ 
&=& \frac14 m_a^2 m_c^2 I^{(0)} \Tr \Big(
\slash p \gamma_\mu \gamma_\lambda \gamma_\rho L \Big) \nonumber\\* 
&\times& \Tr \Big[ \slash k \gamma^\mu \gamma^\lambda \gamma^\rho
\Big( |X'|^2 + |Y'|^2 + X'^*Y' \gamma_5 + Y'^*X'\gamma_5
\Big) \nonumber\\* 
&& + \gamma^\mu \slash k' \gamma^\lambda \gamma^\rho
\Big( |X'|^2 + |Y'|^2 + X'^*Y' \gamma_5 - Y'^*X'\gamma_5\Big) \Big] 
\label{M0M1_1}
\end{eqnarray}
The traces are easily evaluated with the help of the formulas
\begin{eqnarray}
\label{traces4}
\Tr \gamma_\alpha\gamma_\beta\gamma_\gamma\gamma_\delta
& = & 4C_{\alpha\beta\gamma\delta} \nonumber\\*
\Tr \gamma_\alpha\gamma_\beta\gamma_\gamma\gamma_\delta\gamma_5 & = &
-4i \epsilon_{\alpha\beta\gamma\delta} \,,
\end{eqnarray}
where 
\begin{eqnarray}
\label{defC}
C_{\alpha\beta\gamma\delta} & = & 
g_{\alpha\beta}g_{\gamma\delta} - g_{\alpha\gamma}g_{\beta\delta}
+ g_{\alpha\delta}g_{\beta\gamma} 
\end{eqnarray}
When Eq.\ (\ref{traces4}) is used in (\ref{M0M1_1}), four terms are
produced, which we schematically denote as $CC$, $C\epsilon$,
$\epsilon C$ and $\epsilon\epsilon$, indicating which factor $C$ or
$\epsilon$ they contain from each of the two traces that appear. It is
easy to see that the term $CC$ is real, while the terms $C\epsilon$
and $\epsilon C$ are zero after contracting the corresponding Lorentz
indices. Only the term $\epsilon\epsilon$ has a non-zero imaginary
part, and a little bit of algebra shows that
\begin{eqnarray}
\Im \Big< M^{(0)*} M^{(1)}_1 \Big> = 24 m_a^2 m_c^2 I^{(0)} k'
\cdot p \Im (X'^*Y') \,.
\end{eqnarray}
%

\subsection{2nd term}
For the remaining terms, it is useful to use the identity
\begin{eqnarray}
\gamma^\lambda\gamma^\rho\gamma^\mu =
C^{\lambda\rho\mu\alpha}\gamma_\alpha 
+ i\epsilon^{\lambda\rho\mu\alpha}\gamma_\alpha\gamma_5 \,,
\label{3gammas}
\end{eqnarray}
where $C^{\lambda\rho\mu\alpha}$ is defined in Eq.\ (\ref{defC}).  
It then follows that
\begin{eqnarray}
\gamma^\lambda\gamma^\rho\gamma^\mu L =
\Big( C^{\lambda\rho\mu\alpha}
- i\epsilon^{\lambda\rho\mu\alpha} \Big) \gamma_\alpha L \,,
\label{3gammaL}
\end{eqnarray}
and
\begin{eqnarray}
\left< M^{(0)*} M^{(1)}_2 \right>
&=& \frac14 m_a I^{(2)\rho} \Big( C_{\lambda\rho\mu\alpha} -
i\epsilon_{\lambda\rho\mu\alpha} \Big) \nonumber\\*
&& \times \Tr \Big[ (\slash k + m_a) \gamma^\beta (X'^* + Y'^*
\gamma_5) (\slash k' + m_a) \gamma^\lambda \gamma^\mu (X' + Y'
\gamma_5) \Big] \nonumber\\*
&& \times \Tr \Big[ (\slash p + m_b) \gamma_\beta L (\slash p' + m_c)
\gamma^\alpha L\Big] \,.
\label{M0M2}
\end{eqnarray}
It is not difficult to see that all contributions involving the $C$ term 
from the first parenthesis are real.  Among the other terms that are
not necessarily real, some are zero
identically after contracting the Lorentz indices and there are others
that, while not zero identically, are proportional to either one of
the following factors
\begin{eqnarray}
\epsilon^{\lambda\rho\sigma\tau}I^{(2)}_\rho p_\sigma p^\prime_\tau 
k_\lambda\,,
\qquad
\epsilon^{\lambda\rho\sigma\tau}I^{(2)}_\rho p_\sigma p^\prime_\tau 
k^\prime_\lambda \,.
\end{eqnarray}
Since the integral $I^{(2)}_\rho$ is a vector that depends on 
$p^\prime$ and $k^\prime$, it is proportional to either 
$p^\prime_\rho$ or $k^\prime_\rho$. Whence all such terms 
eventually appear contracted in the form
\begin{eqnarray}
\label{epsiloncontraction}
\epsilon^{\lambda\rho\sigma\tau} k_\rho p_\sigma p^\prime_\tau 
k^\prime_\lambda \,,
\end{eqnarray}
and in the end yield zero by momentum conservation.  In summary, in
the first trace that appears in Eq.\ (\ref{M0M2}), only the terms that
contain the combination $X'^*Y' - X'Y'^*$ contribute to the
difference in the rates, and a little bit of algebra yields
\begin{eqnarray}
\label{M0M2final}
\Im \left< M^{(0)*} M^{(1)}_2 \right> = -8 m_a^2 \Im(X'^* Y')
I^{(2)\rho} \Big[ k'_\rho p\cdot p' + p_\rho k' \cdot p' + p'_\rho k'
\cdot p \Big] \,.
\end{eqnarray}
%

\subsection{3rd term}
Using Eq.\ (\ref{3gammas}) once more, 
by straightforward algebraic manipulations we obtain
\begin{eqnarray}
\left< M^{(0)*} M^{(1)}_3 \right>
&=& \frac14 m_c I^{(1)}_\rho \times \Tr \Big[ (\slash p + m_b)
\gamma_\beta L (\slash p' + m_c) \gamma_\lambda \gamma_\mu L \Big]
\nonumber\\* 
&& \times \bigg( C^{\lambda\rho\mu\alpha} \Tr \Big[ (\slash k + m_a)
\gamma^\beta (X'^* + Y'^*\gamma_5) (\slash k' + m_a)
\gamma_\alpha (X' + Y'\gamma_5) \Big] \nonumber\\*
&& + i\epsilon^{\lambda\rho\mu\alpha} \Tr \Big[ (\slash k + m_a)
\gamma^\beta (X'^* + Y'^*\gamma_5) (\slash k' + m_a)
\gamma_\alpha (Y' + X'\gamma_5) \Big] 
\bigg) \,.
\end{eqnarray}
The remaining manipulations and arguments are 
similar to those that lead to Eq.\ (\ref{M0M2final}),
and in this case they yield
%
%
\begin{eqnarray}
\Im \left< M^{(0)*} M^{(1)}_3 \right> = 24 m_a^2 m_c^2 \Im(X'^*Y') 
I^{(1)}_\rho p^\rho \,.
\end{eqnarray}
%

\subsection{4th term}
It is useful to notice that the expression for 
$M^{(1)}_4$ can be simplified by using the identity
in Eq.\ (\ref{3gammas}), and using then the formulas
\begin{eqnarray}
C^{\lambda\rho\mu\alpha}C_{\lambda\nu\mu\beta} & = &
2(\delta^\rho_\nu \delta^\alpha_\beta 
+ \delta^\alpha_\nu \delta^\rho_\beta)
\nonumber\\*
\epsilon^{\lambda\rho\mu\alpha}\epsilon_{\lambda\nu\mu\beta} & = &
- 2(\delta^\rho_\nu \delta^\alpha_\beta 
- \delta^\alpha_\nu \delta^\rho_\beta)\,.
\end{eqnarray}
Thus we obtain
\begin{eqnarray}
\label{M14}
M^{(1)}_4 & = & 4
\Big[\bar u_c(p')\gamma^\alpha L u_b(p) \Big] 
\bigg\{ I^{(12)\beta}_\beta (X' - Y') 
\Big[\bar u_a(k')\gamma_\alpha L u_a(k)\Big]\nonumber\\* 
&&\mbox{} + 
I^{(12)}_{\alpha\beta} (X' + Y')
\Big[\bar u_a(k')\gamma^\beta R u_a(k) \Big] \bigg\} \,,
\end{eqnarray}
and then
\begin{eqnarray}
\left< M^{(0)*} M^{(1)}_4 \right> &=& 
\Tr \left[(\slash p + m_b) \gamma_\mu L(\slash p' + m_c)
\gamma^\alpha L \right] \nonumber\\* 
&\times& \bigg\{
I^{(12)\beta}_\beta (X' - Y')
\Tr \left[ (\slash k + m_a) \gamma^\mu(X'^*
+ Y'^*\gamma_5)(\slash k' + m_a) \gamma_\alpha L \right]
\nonumber\\* 
&& +
I^{(12)}_{\alpha\beta} (X' + Y') \Tr \left[(\slash k + m_a)
\gamma^\mu(X'^* + Y'^*\gamma_5)
(\slash k' + m_a) \gamma^\beta R \right]\bigg\} \,.
\end{eqnarray}
When we carry out the traces and perform the contractions, the terms
that do not have the factor of $m^2_a$ turn out to be real,
proportional either to $|X'-Y'|^2$ or to $|X'+Y'|^2$.  The
only terms that contribute to the rate difference are those that
have the factor of $m^2_a$, and by the same manipulations
that lead to Eq.\ (\ref{M0M2}) we find 
\begin{eqnarray}
\Im \left< M^{(0)*} M^{(1)}_4 \right> = 8m^2_a
\Im (X'^*Y')
I^{(12)}_{\alpha\beta} \Big[p^\alpha p^{\prime\beta} +
p^{\prime\alpha} p^{\beta} + g^{\alpha\beta}p\cdot p' \Big] \,.
\end{eqnarray}
%

\subsection{The sum}
Summarizing the formulas that we have obtained, we can now write
\begin{eqnarray}
\label{Rfinal1}
R = -e^2\Im(X'^* Y^\prime)\left[Z_1 + Z_2 + Z_3 + Z_4\right] \,,
\end{eqnarray}
where
\begin{eqnarray}
\label{Zi}
Z_1 & = & 24 m^2_a m^2_c I^{(0)} p\cdot k' \,, \nonumber\\
Z_2 & = & -8m_a^2
I^{(2)\rho} \Big[ p_\rho(p'\cdot k') + k'_\rho (p\cdot p') +
p'_\rho(p\cdot k') \Big] \,, \nonumber\\
Z_3 & = & 24 m^2_a m^2_c I^{(1)}_\rho p^\rho \,, \nonumber\\
Z_4 & = & 8m^2_a
I^{(12)}_{\alpha\beta} \Big[p^\alpha p^{\prime\beta} +
p^{\prime\alpha} p^{\beta} + g^{\alpha\beta}p\cdot p' \Big] \,.
\end{eqnarray}
It is now obvious that $\bar R$, which is to be computed similarly
but with the substitution indicated in Eq.\ (\ref{M->Mbar}),
is given by $\bar R = -R$, and therefore
\begin{eqnarray}
\label{2R}
\ave{|M|^2} - \ave{|\overline M|^2} =  2R \,.
\end{eqnarray}
Clearly the $Z_i$, and consequently the $CP$ violating effects given
by $R$, vanish in the limit that all the charged lepton masses are
taken to be zero.  We then consider the quantities $Z_i$ evaluated to
the lowest order in the charged lepton masses; i.e., we keep only
those terms that contain two powers of the charged lepton mass. At
this order, $Z_1$ and $Z_3$ do not contribute.  Since $Z_2$ and $Z_4$
already have an explicit factor $m_a^2$, we evaluate the other
kinematic factors, for massless particles.  As shown in
Appendix~\ref{sec:intphasespace}, in this limit the relevant integrals
are given by
\begin{eqnarray}
\label{phasespaceint}
I_\mu^{(2)} & = & B_0 P_\mu - B_1 Q_\mu \nonumber\\
I_{\mu\nu}^{(12)} & = & \frac14 \Big[(B_0 - B_2) sg_{\mu\nu} + 
(2B_0 - B_0 + B_2) P_\mu P_\nu\nonumber\\*
&&\mbox{} +
2 B_1 (Q_\mu P_\nu - P_\mu Q_\nu) - (3B_2 - B_0)Q_\mu Q_\nu \Big]\,, 
\end{eqnarray}
where
\begin{eqnarray}
\label{kinematicdefs}
P & = & k' + p' \,, \nonumber\\
Q & = & k' - p' \,, \nonumber\\
s & = & P^2\,,
\end{eqnarray}
and
\begin{eqnarray}
\label{defBn}
B_n = - {1 \over 16\pi s} 
\int_{-1}^{+1} d\xi \; {\xi^n \over 1-\xi} \,.
\end{eqnarray}
Thus
\begin{eqnarray}
Z_2 & = & -8m^2_a s \left[B_0 p\cdot P - B_1 p\cdot Q\right] \nonumber\\
Z_4 & = & 4m^2_a s \left[ 2B_0 p\cdot P - (B_0 + B_2)p\cdot Q \right] \,, 
\end{eqnarray}
and we finally obtain
\begin{eqnarray}
\label{Z}
Z\equiv Z_2 + Z_4 = 4m^2_a s [2B_1 - B_0 - B_2] \, p\cdot Q +
O(m^4_\ell s)\,. 
\end{eqnarray}
It is reassuring to observe that, while the integrals $B_n$ defined in
Eq.\ (\ref{defBn}) are (infrared) divergent individually, the combination that
appears in Eq.\ (\ref{Z}) is divergent-free, and its value is given by
\begin{eqnarray}
2B_1 - B_0 - B_2 = {1 \over 8\pi s} \,.
\end{eqnarray}
Thus, from Eqs.\ (\ref{Rfinal1}) and (\ref{Z}),
\begin{eqnarray}
R = -\frac{e^2}{2\pi}m_a^2\Im(X'^* Y^\prime)(p\cdot Q) \,,
\end{eqnarray}
and finally, from Eq.\ (\ref{2R}), 
\begin{eqnarray}
\label{ratediff1}
\ave{|M|^2} - \ave{|\overline M|^2} =  
-\frac{e^2}{\pi}m_a^2\Im(X'^* Y^\prime)(p\cdot Q) \,.
\end{eqnarray}

On the other hand, the leading term of the amplitude squared
is straightforward to calculate and yields
\begin{eqnarray}
\ave{|M^{(0)}|^2} = \ave{|\overline M^{(0)}|^2} =  
4|X' + Y'|^2 (k\cdot p')(k'\cdot p) + 
4|X' - Y'|^2 (k\cdot p)(k'\cdot p') \,,
\end{eqnarray}
which determines the total rate. Taking the massless limit approximation,
and using Eq.\ (\ref{ratediff1}), we then have
\begin{eqnarray}
\label{ratediff2}
\frac{1}{\Gamma}\left[
\frac{d\Gamma}{d(\cos\theta)} - \frac{d\bar\Gamma}{d(\cos\theta)}\right] =
-\frac{e^2}{4\pi}
\frac{(m_a^2/s)\Im(X'^* Y^\prime)}
{|X' + Y'|^2 + (1/3) |X' - Y'|^2}\cos\theta \,,
\end{eqnarray}
where $\theta$ is the angle between $\hat k$ and $\hat k'$, in both the 
direct and the conjugate processes.

\section{Discussion and Conclusions}
\label{sec:conclusions}
Using Eq.\ (\ref{XYprime}) and the formulas given in Eqs.\
(\ref{lambdaAB}) and (\ref{lambdaZ}), the $CP$ violating quantity
$\Im(X'^*Y^\prime)$ that appears in Eq.\ (\ref{ratediff2}) is
proportional to
\begin{eqnarray}
\label{delta}
\delta & \equiv & (X + Y) \sum_{i,j,k}
\bigg\{f_A^{(ij)}f^{(k)}_Z \Big[ \Im(t_{bjck})|V_{ai}|^2 +
\Im(t_{aicj}t_{bjck})|V_{cj}|^{-2} \Big] \nonumber\\*
&&\mbox{} + 
f_B^{(ij)}f^{(k)}_Z\Im(s_{aij} s_{bik} s_{ckj})\bigg\}\,,
\end{eqnarray}
where $X,Y$ are the neutral-current couplings defined in 
Eq.\ (\ref{XaYa}),
$f^{(ij)}_{A,B}$ and $f^{(k)}_Z$ are the kinematic factors
defined in Eqs.\ (\ref{fAB}) and (\ref{fZ}) while the coefficients
$t_{bjck}$ and $s_{aij}$ are given by
\begin{eqnarray}
t_{aibj} & = & V_{ai} V_{bj} V^*_{aj} V^*_{bi} \,,\nonumber\\
s_{aij} & = & V_{ai} V^*_{aj} K^*_i K_j \,.
\end{eqnarray}
The $t$ coefficients are the rephasing-invariant parameters of the
lepton sector that are analogous to those introduced for the quark
sector in Refs.\ \cite{Greenberg:1985mr,Dunietz:1985uy}.  These
parameters occur in a purely lepton-number conserving theory.  In
fact, if lepton-number is conserved in the theory, the diagrams in
Fig.\ \ref{f:boxdiagrams}B  do not exist, so that we can put
$f^{(ij)}_B=0$.  In this case, Eq.\ (\ref{delta}) clearly shows that
only the $t$-invariants appear in the CP-violating part of the
amplitude. 

On the other hand, the $s$ coefficients are precisely the rephasing
invariants introduced in Ref. \cite{Nieves:1987pp} to accommodate the
Majorana neutrinos. As shown in there, and further
studied in Ref.\ \cite{Nieves:2001fc}, the $s$ coefficients form a
suitable set of rephasing-invariant parameters for describing the $CP$
violating effects due to the Majorana nature of the neutrinos.  In
fact, the dependence of $\delta$ on the product of three $s$
parameters, as indicated in Eq.\ (\ref{delta}), was anticipated in
Ref.\ \cite{Nieves:2001fc} [e.g., Eqs.\ (3.27) and (3.37) of that
paper].

Thus, the present calculation confirms the expectation
that the extra $CP$ violating phases that exist for Majorana neutrinos
can appear in $CP$ violating observables
in processes that conserve total lepton number. Although
the type of process that we have specifically considered
(e.g., $e + \mu \rightarrow e + \tau$) is not a realistic one at present,
related processes such as $\tau \rightarrow \bar e + e + \mu$
will show the same effect. The corresponding calculations for the latter
kind of process is more involved than those presented here due to the
three-body final state involved. Nevertheless, 
the present calculations, besides serving as a proof of concept, 
set the stage for considering such three-body decay process,
and should prove to be technically useful in that context as well.

\paragraph*{Acknowledgments~: }
PBP wants to thank Gautam Bhattacharyya for numerous discussions.  The
work of JFN has been partially supported by the U.S.\  National Science
Foundation Grant No. PHY-0139538.

\appendix
\section{Calculation of the box diagrams}
\label{sec:calcboxdiagrams}
We calculate here the diagrams shown in Fig.\ \ref{f:boxdiagrams}, 
for arbitrary incoming and outgoing lepton flavors. We denote
the momentum vectors by $k_\ell$, with $\ell = a,b,c,d$ according
to the labels assigned in the diagrams.

\subsection*{Diagram (\ref{f:boxdiagrams}A)}
Not counting the exchange diagrams, we have four diagrams
altogether. However, those that involve either one or two unphysical
Higgs in the internal lines are of order $1/M_W^6$, and we neglect them.  
Therefore, in
terms of the couplings of the charged current $j_\mu^{(W)} =
\sum\limits_{a,i}V_{a i} \bar\ell_a\gamma_\mu L\nu_i$,
\begin{eqnarray}
\label{Mbdef}
iM^{(abcd)}_A & = & \left(\frac{-ig}{\sqrt{2}}\right)^4
\sum_{i,j}\int\frac{d^4q}{(2\pi)^4}
\left[V_{d j}V^*_{b j}\bar u_d(k_d)\gamma_\rho LiS_{\nu_j}(k_d - k_a + q)
\gamma_\nu L u_b(k_b)\right]
\nonumber\\*
& \times &
\left[V_{c i}V^*_{a i}\bar u_c(k_c)\gamma_\mu LiS_{\nu_i}(q)
\gamma_\lambda L u_a(k_a)\right]
\left(\frac{-ig^{\mu\nu}}{(k_c - q)^2 - M_W^2}\right)
\left(\frac{-ig^{\lambda\rho}}{(k_a - q)^2 - M_W^2}\right)
\,.\nonumber\\* 
\end{eqnarray}
This expression can be written in the form
\begin{eqnarray}
iM^{(abcd)}_A = \left(\frac{g}{\sqrt{2}}\right)^4
\sum_{i,j} (V_{d j}V^*_{b j}V_{c i}V^*_{a i})
I^\rho_\lambda
\left[\bar u_d(k_d)\gamma_\nu\gamma_\rho\gamma_\mu L u_b(k_b)
\vphantom{a^i}\right]
\left[\bar u_c(k_c)\gamma^\mu\gamma^\lambda\gamma^\nu L u_a(k_a)\right] \,,
\end{eqnarray}
where
\begin{eqnarray}
I^\rho_\lambda \equiv \int\frac{d^4 q}{(2\pi)^4}\frac{q^\rho q_\lambda}
{[(k_c - q)^2 - M_W^2] [(k_a - q)^2 - M_W^2][q^2 - m_{\nu_i}^2] 
[(k_d - k_a + q)^2 - m_{\nu_j}^2]} \,.
\end{eqnarray}
Neglecting terms $O(1/M_W^6)$,
\begin{eqnarray}
I_{\rho\lambda} = \frac{1}{4}g_{\rho\lambda} 
\left(\frac{i}{16\pi^2 M_W^2}\right)f^{(ij)}_A
\end{eqnarray}
where $f^{(ij)}_A$ is given in Eq.\ (\ref{fAB}) in the text.
Finally, using the identity
\begin{eqnarray}
\left[\bar u_d(k_d)\gamma_\nu\gamma_\lambda\gamma_\mu L u_b(k_b)
\vphantom{a^i}\right]
\left[\bar u_c(k_c)\gamma^\mu\gamma^\lambda\gamma^\nu L u_a(k_a)\right] = 
16 
\left[\bar u_d(k_d)\gamma_\lambda L u_b(k_b)\right]
\left[\bar u_c(k_c)\gamma^\lambda L u_a(k_a)\right] \,, \nonumber\\*
\end{eqnarray}
we arrive at
\begin{equation}
M^{(abcd)}_A  = 
\frac{g^4}{64\pi^2 M_W^2} 
\sum_{i,j} (V^*_{a i}V^*_{b j}V_{c i}V_{d j})(4f^{(ij)}_A){\cal
M}^{(abcd)}_W 
\end{equation}
where
\begin{equation}
\label{calMW}
{\cal M}^{(abcd)}_W \equiv
[\bar u_d(k_d)\gamma^\mu L u_b(k_b)][\bar u_c(k_c)\gamma_\mu L
u_a(k_a)] \,. 
\end{equation}
The contribution to the physical amplitude, including the exchange term, 
is given by
\begin{equation}
M_A(\scatt abcd) = M^{(abcd)}_A - M^{(abdc)}_A \,.
\end{equation}
Using the fact that ${\cal M}^{(abcd)}_W = -{\cal M}^{(abdc)}_W$,
which follows from a Fierz transformation, the contribution to the
physical amplitude is then
\begin{equation}
M_A(\scatt abcd)  = \lambda^{(abcd)}_A {\cal M}^{(abcd)}_W \,,
\end{equation}
with $\lambda^{(abcd)}_A$ as defined in Eq.\ (\ref{lambdaAB}) in the text.

\subsection*{Diagram (\ref{f:boxdiagrams}B)}
As with the diagrams \ref{f:boxdiagrams}A,
there are four diagrams not counting the exchange diagrams, 
and those that
involve either one or two unphysical Higgs in the internal lines are
of order $1/M_W^6$.  Therefore, to order $1/M^4_W$,
\begin{eqnarray}
\label{Mddef}
iM^{(abcd)}_B & = & \left(\frac{g}{\sqrt{2}}\right)^4
\sum_{i,j}\int\frac{d^4q}{(2\pi)^4}\left[
\bar u_c(k_c)(-i\gamma_\nu LV_{c j})
iK_j^{*\,2}S_{\nu_j}(k_a - k_d - q)
(i\gamma_\rho RV_{d j})v_d(k_d)\right]\nonumber\\*
& \times &
\left[\bar v_b(k_b)(i\gamma_\mu R V^*_{b i})i K^2_i S_{\nu_i}(q)
(-i\gamma_\lambda LV^*_{a i}) u_a(k_a)\right]
\left[\frac{-ig^{\mu\nu}}{(k_b + q)^2 - M_W^2}\right]
\left[\frac{-ig^{\lambda\rho}}{(k_a - q)^2 - M_W^2}\right]\,,\nonumber\\* 
\end{eqnarray}
where we have used the Majorana condition of Eq.\ (\ref{Majocond}).
Only the mass terms of the neutrino propagators contribute due to the
$L$ and $R$ factors on opposite sides of the propagator, and we obtain
\begin{eqnarray}
\label{Mdform}
iM^{(abcd)}_B & = & \left(\frac{g}{\sqrt{2}}\right)^4
\sum_{i,j}\left(m_{\nu_i}m_{\nu_j}K_i^2 K_j^{*\,2}
V_{c j} V_{dj} V_{b i}^* V_{a i}^*\right)iI_B \nonumber\\*
&& \times
\Big[ \bar u_c(k_c)\gamma^\lambda\gamma^\rho R v_d(k_d) \Big]
\Big[ \bar v_b(k_b)\gamma_\lambda\gamma_\rho L u_a(k_a) \Big] \,,
\end{eqnarray}
where
\begin{eqnarray}
iI_B \equiv \int\frac{d^4 q}{(2\pi)^4}\frac{1}
{[(k_b + q)^2 - M_W^2][(k_a - q)^2 - M_W^2][q^2 - m_{\nu_i}^2]
[(k_a - k_d - q)^2 - m_{\nu_j}^2]} \,.
\end{eqnarray}
Neglecting terms $O(1/M_W^6)$, the explicit evaluation of $I_B$
yields 
\begin{eqnarray}
I_B = \frac{1}{16\pi^2 M_W^2} 
\left\{
\frac{r_i \log r_i}{r_j - r_i} - r_i + (i\leftrightarrow j)
\right\} \,.
\end{eqnarray}
Finally, by a Fierz transformation, 
\begin{eqnarray}
\label{MdFierzed}
\Big[ \bar u_c(k_c)\gamma^\lambda\gamma^\rho R v_d(k_d) \Big]
\Big[ \bar v_b(k_b)\gamma_\lambda\gamma_\rho L u_a(k_a) \Big] = 
2\Big[ \bar u_d(k_d)\gamma^\lambda L u_b(k_b) \Big]
\Big[ \bar u_c(k_c)\gamma_\lambda L u_a(k_a) \Big] \,,
\end{eqnarray}
the details of which are shown in Appendix~\ref{sec:fierz}. 
Thus Eq.\ (\ref{Mdform}) reduces to
\begin{equation}
M^{(abcd)}_B = 
\frac{g^4}{64\pi^2 M_W^2} 
\sum_{i,j}\left(K_i^2 K_j^{*\,2} 
V_{ai}^* V_{bi}^* V_{cj} V_{dj}\right)(2f^{(ij)}_B) {\cal M}^{(abcd)}_W\,,
\end{equation}
where ${\cal M}^{(abcd)}_W$ has been defined in Eq.\ (\ref{calMW}).
As with the previous diagram, the contribution to the physical
amplitude, including the exchange term, is given by
\begin{eqnarray}
M_B(\scatt abcd) & = & M^{(abcd)}_B - M^{(abdc)}_B \nonumber\\
& = & \lambda^{(abcd)}_B {\cal M}^{(abcd)}_W \,,
\end{eqnarray}
with $\lambda^{(abcd)}_B$ as defined in Eq.\ (\ref{lambdaAB}).

\section{Fierz transformations}
\label{sec:fierz}
Fierz transformations concern products of two fermion bilinears.  For
arbitrary spinors $w_1$, $w_2$, $w_3$ and $w_4$, we denote these
products collectively as $e_i$ with $i = S, V, T, A, P$.  In other
words,
\begin{eqnarray}
e_i = [\bar w_1 \Gamma^i w_2][\bar w_3 \Gamma_i w_4] \,,
\label{ei}
\end{eqnarray}
with
\begin{eqnarray}
\Gamma_i = (1,\gamma_\mu,\sigma_{\mu\nu},\gamma_\mu\gamma_5,\gamma_5)
\,, 
\end{eqnarray}
and $\Gamma^i$ being the same things with contravariant Lorentz indices.
Then we will denote by $e'_i$ the same quantities, but with 
$w_2 \leftrightarrow w_4$; i.e., 
\begin{eqnarray}
e'_i = [\bar w_1\Gamma^i w_4][\bar w_3\Gamma_i w_2] \,.
\end{eqnarray}
The basic Fierz identity is the relation between the two sets of
bilinears, $\{e_i\}$ and $\{e'_i\}$,
\begin{eqnarray}
e_i = \sum_j F_{ij}e'_j \,,
\end{eqnarray}
where
\begin{eqnarray}
F = 
\frac{1}{4}\left(\begin{array}{rrrrr}
1 & 1 & \frac{1}{2} & -1 & 1 \\
4 & -2 & 0 & -2 & -4 \\
12 & 0 & -2 & 0 & 12\\
-4 & -2 & 0 & -2 & 4\\
1 & -1 & \frac{1}{2} & 1 & 1
\end{array}\right) \,.
\end{eqnarray}

To prove Eq.\ (\ref{MdFierzed}) from here, first note that the
identity 
\begin{eqnarray}
\gamma^\lambda\gamma^\rho = g^{\lambda\rho} - i\sigma^{\lambda\rho} 
\end{eqnarray}
can be used to write the left hand side of Eq.\ (\ref{MdFierzed}) as
$4e_S - e_T$ in the notation of Eq.\ (\ref{ei}), 
with
\begin{eqnarray} 
w_1 & = & u_{Lc}(k_c) \nonumber\\
w_2 & = & v_{Rd}(k_d) \nonumber\\
w_3 & = & v_{Rb}(k_b) \nonumber\\
w_4 & = & u_{La}(k_a) \,.
\end{eqnarray}
The relevant Fierz formula for us now is
\begin{eqnarray}
4e_S - e_T & = &\left(e'_S + e'_P + \frac{1}{2}e'_T
+ e'_V - e'_A\right) - \left(3e'_S + 3e'_P
- \frac{1}{2}e'_T\right) \nonumber\\
& = & -2(e'_S + e'_P) + e'_T + e'_V - e'_A \,.
\end{eqnarray}
Substituting the spinors $a$-$d$ given above, 
it turns out that $e'_{S,P,T} = 0$, while
\begin{eqnarray}
\mbox{} -e'_A = e'_V & = & 
\Big[ \bar u_{Lc}(k_c)\gamma^\lambda u_{La}(k_a) \Big]
\Big[ \bar v_{Rb}(k_b)\gamma_\lambda v_{Rd}(k_d) \Big] \nonumber\\
& = & \Big[ \bar u_{Lc}(k_c)\gamma^\lambda u_{La}(k_a) \Big]
\Big[ \bar u_{Ld}(k_d)\gamma_\lambda u_{Lb}(k_b) \Big] \,,
\end{eqnarray}
using Eq.\ (\ref{ccvector}) and similar relations in the last step.
This gives Eq.\ (\ref{MdFierzed}).

\section{Derivation of Eq.\ (\ref{M1a})}
\label{sec:cutkosky}
In order to make the discussion generally applicable and not tied
to any particular channel, we consider the process labeled as
\begin{eqnarray}
\ell_a(k_a) + \ell_b(k_b) \rightarrow \ell_c(k_c)
+ \ell_d(k_d) \,.
\end{eqnarray}
Thus, this includes processes such as $\scatt abaa$ and $\scatt
abac$. The treatment for the crossed processes such as $\bscatt abaa$
is similar, with the appropriate modifications dictated by the usual
substitution (crossing) rules.  The amplitude $M^{(0)}$, determined
from the diagrams that are schematically represented in Fig.\
\ref{f:1-loop}, can be expressed in the form
\begin{equation}
\label{schematicM0}
iM^{(0)} = i\sum_{A,B} g_{AB}I_{AB}(k_a,k_b,k_c,k_d)
\Big[\bar u_c(k_c)\Gamma_A u_a(k_a) \Big]
\Big[\bar u_d(k_d)\Gamma_B u_b(k_b) \Big]\,,
\end{equation}
where
each $g_{AB}$ denotes a product of coupling constants, each $I_{AB}$
is Feynman integral which is real, and the $\Gamma_A$ are the
generalized Dirac-Pauli matrices. In the most general case, the
amplitude can be brought to this form by making the appropriate Fierz
transformations.  With this in place, the amplitude for $M^{(1)}$,
determined from the diagrams represented in Fig.~\ref{f:2-loop}, is
given by
\begin{eqnarray}
\label{M1-2loop}
iM^{(1)} & = & \int\frac{d^4q}{(2\pi)^4} i\sum_{A,B}
g_{AB}I_{AB}(k_a,k_b,k_c + q,k_d - q) iD_F^{\nu\mu}(q) 
\nonumber\\*
&&\times \left[\bar u_c(k_c)(-ie_\gamma\gamma_\mu)iS_{Fc}(k_c + q)
\Gamma_A u_a(k_a)\right]
\nonumber\\*
&&\times \left[\bar u_d(k_d)(-ie_d\gamma_\nu)iS_{Fd}(k_d - q)
\Gamma_B u_b(k_b)\right]\,.
\end{eqnarray}
Our task is to determine the absorptive part of $M^{(1)}$.  

Since the integrals $I_{AB}$ are real, the absorptive part can
arise only from the denominators of the lepton propagators
in Eq.\ (\ref{M1-2loop}), which we write in the form
\begin{equation}
D = (q^0 - a)(q^0 - a')(q^0 - b)(q^0 - b') \,,
\end{equation}
where
\begin{eqnarray}
\label{M1poles}
a\phantom{'} & = & E^{(c)}_{k_c + q} - E^{(c)}_{k_c} 
- i\epsilon \nonumber\\
a' & = & -(E^{(c)}_{k_c + q} + E^{(c)}_{k_c}) + i\epsilon \nonumber\\
b\phantom{'} & = & E^{(d)}_{k_d} + E^{(d)}_{k_d - q} 
- i\epsilon \nonumber\\
b' & = & E^{(d)}_{k_d} - E^{(d)}_{k_d - q} + i\epsilon \,,
\end{eqnarray}
with
\begin{eqnarray}
E^{(\ell)}_p = \sqrt{\vec p^{\;^2} + m_\ell^2} \,.
\end{eqnarray}
Therefore, we rewrite Eq.\ (\ref{M1-2loop}) as
\begin{equation}
\label{M1structure}
iM^{(1)} = \int\frac{d^4q}{(2\pi)^4}\frac{1}{D}{i\mathcal M}^{(1)} \,,
\end{equation}
where
\begin{eqnarray}
\label{calM1}
{i\mathcal M}^{(1)} & = & 
i\sum_{A,B}
g_{AB}I_{AB}(k_a,k_b,k_c + q,k_d - q) iD_F^{\nu\mu}(q) 
\nonumber\\*
&&\times \left[\bar u_c(k_c)(-ie_c\gamma_\mu)
i(\slash k_c + \slash q + m_c) \Gamma_A u_a(k_a)\right]
\nonumber\\*
&&\times \left[\bar u_d(k_d)(-ie_d\gamma_\nu)
i(\slash k_d - \slash q + m_d) \Gamma_B u_b(k_b)\right]\,.
\end{eqnarray}

\begin{figure}
\begin{center}
\begin{picture}(200,150)(-100,-75)
\Text(-50,-10)[t]{$C$}
\ArrowLine(-100,0)(0,0)
\ArrowLine(0,0)(100,0)
\Line(0,-75)(0,75)
\Text(-50,30)[b]{$a'$}
\Text(20,30)[b]{$b'$}
\Text(40,-40)[b]{$a$}
\Text(70,-40)[b]{$b$}
\Vertex(-50,20){1}
\Vertex(20,20){1}
\Vertex(40,-30){1}
\Vertex(70,-30){1}
\end{picture}
\end{center}
\caption[]{\label{fig:origpath}
Original path of integration in the $q^0$ plane.}
\end{figure}
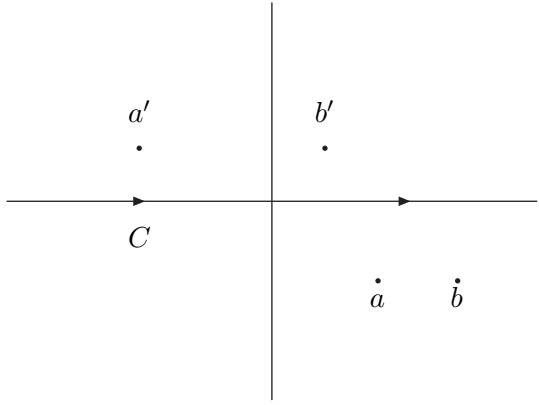

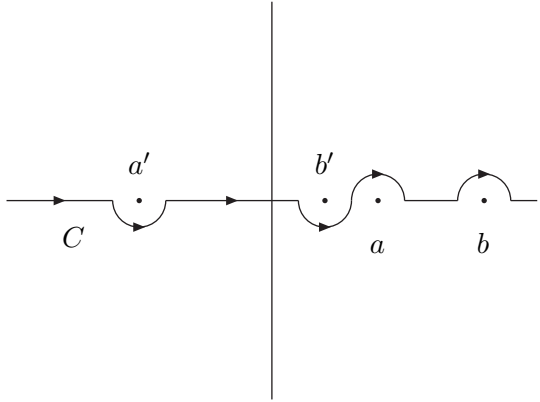
\begin{figure}
\begin{center}
\begin{picture}(200,150)(-100,-75)
\Text(-75,-10)[t]{$C$}
\ArrowLine(-100,0)(-60,0)
\ArrowLine(-40,0)(10,0)
\Line(50,0)(70,0)
\Line(90,0)(100,0)
\ArrowArc(-50,0)(10,180,0)
\ArrowArc(20,0)(10,180,0)
\ArrowArcn(40,0)(10,180,0)
\ArrowArcn(80,0)(10,180,0)
\Line(0,-75)(0,75)
\Text(-50,10)[b]{$a'$}
\Text(20,10)[b]{$b'$}
\Text(40,-20)[b]{$a$}
\Text(80,-20)[b]{$b$}
\Vertex(-50,0){1}
\Vertex(20,0){1}
\Vertex(40,0){1}
\Vertex(80,0){1}
\end{picture}
\end{center}
\caption[]{
\label{fig:avoidpath}
Deformed path of integration in the $q^0$ plane to avoid the poles
when $\epsilon\rightarrow 0$.
}
\end{figure}

{}From a mathematical point of view, the absorptive part of 
$M^{(1)}$ arises from the fact that, as a function of $q^0$,
the integrand in Eq.\ (\ref{M1structure}) has poles at the
points indicated in Eq.\ (\ref{M1poles}), and illustrated
schematically in Fig.\ \ref{fig:origpath}. 
%
%
As $\epsilon\rightarrow 0$ those poles lie on the real axis but,
as long as the path of integration
can be deformed such that it avoids the poles, the resulting
integral is real and the absorptive part of $M^{(1)}$ is zero.
This is what happens for all the kinematic configurations
in which none of the poles coincide with another one, as
illustrated in Fig.\ \ref{fig:avoidpath}.
%
%
However, if the kinematic configuration is such that, in the limit
$\epsilon\rightarrow 0$, one of the poles that lie above the real axis
coincide with one that lies below the real axis, then the path of
integration is \emph{pinched} between the two points, and it cannot be
deformed to avoid the poles. In this case the amplitude will develop
an absorptive part. This side of the coin is also illustrated in Fig.\
\ref{fig:avoidpath}, if we consider the case that $b' = a$. As a
matter of fact, from the condition that $E^{(\ell)}_p$ is a positive
quantity, it follows from Eq.\ (\ref{M1poles}) that this is the only
possible \emph{pinch condition}.

\begin{figure}
\begin{center}
\begin{picture}(200,175)(-100,-100)
\Line(-100,0)(100,0)
\Line(0,-100)(0,75)
\DashArrowLine(-100,-70)(50,-70){4}
\DashArrowLine(50,-70)(50,-10){4}
\DashArrowLine(50,-10)(100,-10){4}
\Text(-25,-80)[t]{$C_2$}
\ArrowArcn(25,-30)(15,0,180)
\ArrowArcn(25,-30)(15,180,0)
\Text(25,-50)[t]{$C_1$}
\Text(-50,30)[b]{$a'$}
\Text(25,30)[b]{$b'$}
\Text(25,-40)[b]{$a$}
\Text(70,-40)[b]{$b$}
\Vertex(-50,20){1}
\Vertex(25,20){1}
\Vertex(25,-30){1}
\Vertex(70,-30){1}
\end{picture}
\end{center}
\caption[]{
\label{fig:deformpath}
Deformed path of integration in the $q^0$ plane.
}
\end{figure}
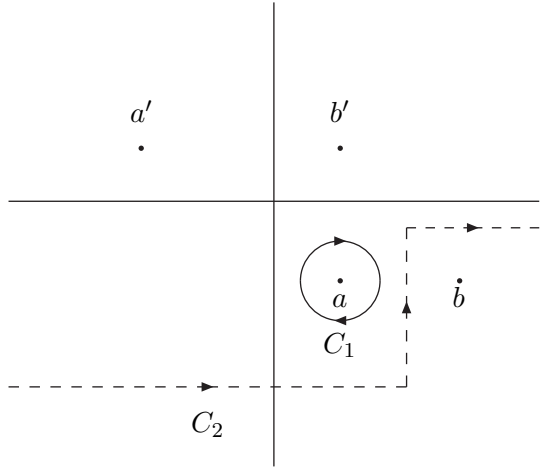

In order to isolate the contribution to $M^{(1)}$ in this situation
we proceed as follows. By Cauchy's theorem, we can deform
the original path of integration as shown in Fig.\ \ref{fig:deformpath}.
%
%
The virtue of this is that we can write
\begin{eqnarray}
\label{M1split}
iM^{(1)} = \left(\int\frac{d^4q}{(2\pi)^4}\frac{1}{D}{i\mathcal M}^{(1)}
\right)_{C_1} + 
\left(\int\frac{d^4q}{(2\pi)^4}\frac{1}{D}{i\mathcal M}^{(1)}\right)_{C_2}\,.
\end{eqnarray}
By the same argument that we have explained above, it now follows that
the integral over the path $C_2$ does not produce an absorptive part
because the path cannot be pinched (i.e., there is no kinematic
configuration for which $b$ can become equal to $a'$ or 
$b'$). Therefore, the integral over $C_2$ contributes only
to the dispersive part of the amplitude and we neglect it.
The integral over $C_1$ on the other hand, can be evaluated
by the method of residues, and therefore
\begin{eqnarray}
iM^{(1)} = -i\int\frac{d^3q}{(2\pi)^3}\frac{1}{D'}{i\mathcal M}^{(1)}\,,
\end{eqnarray}
where
\begin{eqnarray}
D' = 2E^{(c)}_{k_c + q}
\left[E^{(c)}_{k_c} + E^{(d)}_{k_d} - E^{(c)}_{k_c + q} + 
E^{(d)}_{k_d - q}\right]
\left[E^{(c)}_{k_c}  + E^{(d)}_{k_d} - E^{(c)}_{k_c + q} - 
E^{(d)}_{k_d - q} + i\epsilon\right] \,.
\end{eqnarray}
Once again, this has a dispersive and an absorptive contribution.
Retaining only the latter, which is obtained by the substitution
$1/(x + i\epsilon)\rightarrow -i\pi\delta(x)$, we finally obtain
\begin{eqnarray}
\label{M1abs}
iM^{(1)} = -\pi\int\frac{d^3q}{(2\pi)^3}
\frac{1}{2E^{(c)}_{k_c + q}}\frac{1}{2E^{(d)}_{k_d - q}}
\delta(E^{(c)}_{k_c} + E^{(d)}_{k_d}  - 
E^{(c)}_{k_c + q} - E^{(d)}_{k_d - q}) {i\mathcal M}^{(1)}\,.
\end{eqnarray}
To write this in its final form, we put in the expression for 
${\mathcal M}^{(1)}$ 
\begin{eqnarray}
\vec q_c & \equiv & \vec k_c + \vec q \,,\nonumber\\
\vec q_d & \equiv & \vec k_d - \vec q \,,
\end{eqnarray}
and insert the factor 
\begin{eqnarray}
\int\frac{d^3q_c}{(2\pi)^3}(2\pi)^3
\delta^{(3)}(\vec q_c - \vec k_c - \vec q)
\int\frac{d^3q_d}{(2\pi)^3}(2\pi)^3
\delta^{(3)}(\vec q_d - \vec k_d + \vec q) \,.
\end{eqnarray}
When these are substituted in Eq.\ (\ref{M1abs}), the integral over
$d^3\vec q$ can be eliminated with the help of the delta functions,
and we finally arrive at
\begin{eqnarray}
\label{M1absfinal}
iM^{(1)} & = & -\left(\frac{1}{2}\right)
\int\frac{d^3q_c}{(2\pi)^3 2E^{(c)}_{q_c}}
\int\frac{d^3q_d}{(2\pi)^3 2E^{(d)}_{q_d}}
(2\pi)^4\delta(q_c + q_d - k_c - k_d)
i{\mathcal M}^{(1)}\,,
\end{eqnarray}
with ${\mathcal M}^{(1)}$ expressed in the form
\begin{eqnarray}
i{\mathcal M}^{(1)} & = &
i\sum_{A,B}
\left(g_{AB}I_{AB}(k_a,k_b,q_c,q_d)\right) 
iD_F^{\nu\mu}(q_c - k_c) \nonumber\\*
&&\times \left[\bar u_c(k_c)(-ie_c\gamma_\mu)
i(\slash q_c + m_c) \Gamma_A u_a(k_a)\right]
\nonumber\\*
&&\times \left[\bar u_d(k_d)(-ie_d\gamma_\nu)
i(\slash q_d + m_d) \Gamma_B u_b(k_b)\right]\,.
\end{eqnarray}
Using the relation
\begin{eqnarray}
(\slash q_\ell + m_\ell) = \sum_{s} u_\ell(q_\ell)\bar u_{\ell}(q_\ell) \,,
\end{eqnarray}
it is easily seen that the result given in Eq.\ (\ref{M1absfinal})
is equivalent to the formula quoted in Eq.\ (\ref{M1a}).

\section{Integrals over intermediate states}
\label{sec:intphasespace}
Here we consider the evaluation of the integrals defined in Eq.\
(\ref{I...}), the results of which are quoted in Eq.\
(\ref{phasespaceint}).  For the reasons mentioned in the text, we take
all the lepton masses to be zero.

The measure $dL$, defined in Eq.\ (\ref{dX}), contains the factor
\begin{eqnarray}
(k'-q_1)^2 = - \frac12 s (1-\xi) \,,
\end{eqnarray}
where $\xi$ is the cosine of the angle between $\vec q_1$ and $\vec
k'$, and $s$ has been defined in Eq.\ (\ref{kinematicdefs}).
Then, for any integrand $F$, we can write
\begin{eqnarray}
\int dL\; F = - {1 \over 8\pi s} \int_{-1}^{+1} d\xi \;
{F \over 1- \xi} \,,
\end{eqnarray}
performing as many integrations as the delta function allows us.
Further, complementing Eq.\ (\ref{kinematicdefs}) it is convenient to
define
\begin{eqnarray}
P' & = & q_1 + q_2 \nonumber\\*
Q' & = & q_1 - q_2 \,.
\end{eqnarray}
Noting that momentum conservation ensures that $P'_\mu = P_\mu$,
it follows that
\begin{eqnarray}
I_\mu^{(1)} = \frac12 \int dL\;(P'_\mu + Q'_\mu) = 
\frac12 \left( P_\mu J + J_\mu \right) \,,
\label{I1}
\end{eqnarray}
and similarly
\begin{eqnarray}
I_\mu^{(2)} = \frac12 \left( P_\mu J - J_\mu \right) \,,
\label{I2}
\end{eqnarray}
where we define a new set of integrals
\begin{eqnarray}
J & = & \int dL\nonumber\\
J_{\mu_1\mu_2 \cdots \mu_n} & \equiv &
\int dL \; Q'_{\mu_1} Q'_{\mu_2} \cdots Q'_{\mu_n}  \,.
\label{Jn}
\end{eqnarray}
For the integral with two indices, we can similarly write
\begin{eqnarray}
\label{I12J}
I_{\mu\nu}^{(12)} &=& \frac14 \left( P_\mu P_\nu J + 
J_\mu P_\nu - P_\mu J_\nu - J_{\mu\nu}\right) \,.
\end{eqnarray}
Therefore, we only have to evaluate the integrals in Eq.\ (\ref{Jn}).

The scalar integral $J$ can be determined immediately,
\begin{eqnarray}
J &=& - {1 \over 8\pi s} \int_{-1}^{+1} d\xi {1 \over 1-\xi} 
= 2B_0 \,,
\label{J}
\end{eqnarray}
where $B_n$ is defined in Eq.\ (\ref{defBn}).  For the others, notice
that
\begin{eqnarray}
P^{\mu_i} J_{\mu_1\mu_2 \cdots \mu_n} = 0
\label{P.J}
\end{eqnarray}
for any $i=1,2,\cdots n$.  For the one-index integral, there is
only one such relation
\begin{eqnarray}
P^\mu J_\mu = 0 \,,
\end{eqnarray}
which implies that 
\begin{eqnarray}
J_\mu = b Q_\mu \,,
\end{eqnarray}
for some invariant $b$.  The invariant can be determined by
contracting both sides with $Q^\mu$, which gives
\begin{eqnarray}
b = {Q^\mu J_\mu \over Q^2} \,.
\end{eqnarray}
By explicit computation,
\begin{eqnarray}
Q^\mu J_\mu = - 2s B_1 \,,
\end{eqnarray}
and using $Q^2 = -s$, we arrive at
\begin{eqnarray}
J_\mu = 2B_1 Q_\mu \,.
\end{eqnarray}
Substituting this in Eq.\ (\ref{I2}) then yields
the formula for $I_\mu^{(2)}$ quoted in 
Eq.\ (\ref{phasespaceint}). An analogous formula for $I_\mu^{(1)}$
follows from Eq.\ (\ref{I1}).

For the two-index integral $J_{\mu\nu}$ the relation
in Eq.\ (\ref{P.J}) now dictates the general form
\begin{eqnarray}
J_{\mu\nu} = \lambda(P_\mu P_\nu - sg_{\mu\nu}) + 
\rho Q_\mu Q_\nu \,,
\end{eqnarray}
with the coefficients being easily determined by contracting with
$g^{\mu\nu}$ and by $Q^\mu Q^\nu$.  The equations obtained
this way are
\begin{eqnarray}
s^2 \lambda + s^2 \rho &=& Q^\mu Q^\nu J_{\mu\nu} \nonumber\\*
3s \lambda + s \rho &=& - g^{\mu\nu} J_{\mu\nu} \,.
\label{J2eqns}
\end{eqnarray}
The combinations on the right side of these equations are computed
explicitly as
\begin{eqnarray}
g^{\mu\nu} J_{\mu\nu} &=& \int dL \; Q'^2 =
-2s B_0 \nonumber\\*
Q^\mu Q^\nu J_{\mu\nu} &=& \int dL \; (Q\cdot Q')^2 = 
2s^2 B_2 \,.
\end{eqnarray}
Substituting these in Eq.\ (\ref{J2eqns}) and
solving for $\lambda$ and $\rho$, 
\begin{eqnarray}
J_{\mu\nu} = 
(B_0 - B_2) \Big( P_\mu P_\nu - sg_{\mu\nu} \Big)
+ (3B_2 - B_0)Q_\mu Q_\nu \,,
\end{eqnarray}
which together with Eq.\ (\ref{I12J}) yields the formula for
$I_{\mu\nu}^{(12)}$ quoted in Eq.\ (\ref{phasespaceint}).

%
%

\end{document}